\documentclass[10pt,twocolumn,conf]{IEEEtran}
\usepackage{times}
\usepackage{amsbsy}
\usepackage{latexsym}
\usepackage{amsmath}
\usepackage[ansinew]{inputenc}
\usepackage[dvips ]{graphicx}
\input epsf
\usepackage{epic,eepic}
\usepackage{url}
\usepackage{algorithm}
\usepackage{color}
\usepackage{booktabs}
\usepackage{setspace}
\usepackage{cite}
\IEEEoverridecommandlockouts

\title{\Huge Massive MIMO Systems: Signal Processing Challenges and Research Trends \vspace{0.05em}}
\author{Rodrigo C. de Lamare  \\ Centre for Telecommunications Studies (CETUC) \\ Pontifical Catholic University of Rio de Janeiro, Gávea - 22453-900, Rio de Janeiro, Brazil \\ Communications Research Group \\ Department of Electronics,
    University of York, York Y010 5DD, United Kingdom \\
    Email: \protect\url{delamare@cetuc.puc-rio.br}
\thanks{\footnotesize The work of the author is
supported by the Pontifical Catholic University of Rio de Janeiro
and by the University of York, York Y010 5DD, United Kingdom.   }}

\begin{document}
\maketitle

\begin{abstract}

This article presents a tutorial on multiuser multiple-antenna
wireless systems with a very large number of antennas, known as
massive multi-input multi-output (MIMO) systems. Signal processing
challenges and future trends in the area of massive MIMO systems are
presented and key application scenarios are detailed. A linear
algebra approach is considered for the description of the system and
data models of massive MIMO architectures. The operational
requirements of massive MIMO systems are discussed along with their
operation in time-division duplexing mode, resource allocation and
calibration requirements. In particular, transmit and receiver
processing algorithms are examined in light of the specific needs of
massive MIMO systems. Simulation results illustrate the performance
of transmit and receive processing algorithms under scenarios of
interest. Key problems are discussed and future trends in the area
of massive MIMO systems are pointed out.

\end{abstract}

\section{Introduction}

Wireless networks are experiencing a very substantial increase in
the delivered amount of data due to a number of emerging
applications that include machine-to-machine communications and
video streaming \cite{cisco,lte-a,wlan}. This very large amount of
data exchange is expected to continue and rise in the next decade or
so, presenting a very significant challenge to designers of wireless
communications systems. This constitutes a major problem, not only
in terms of exploitation of available spectrum resources, but also
regarding the energy efficiency in the transmission and processing
of each information unit (bit) that has to substantially improve.
The Wireless Internet of the Future (WIoF) will have therefore to
rely on technologies that can offer a substantial increase in
transmission capacity as measured in bits/Hz but do not require
increased spectrum bandwidth or energy consumption.

Multiple-antenna or multi-input multi-output (MIMO) wireless
communication devices that employ antenna arrays with a very large
number of antenna elements which are known as massive MIMO systems
have the potential to overcome those challenges and deliver the
required data rates, representing a key enabling technology for the
WIoF \cite{marzetta_first}-\cite{nam}. Among the devices of massive
MIMO networks are user terminals, tablets, and base stations which
could be equipped with a number of antenna elements with orders of
magnitude higher than current devices. Massive MIMO networks will be
structured by the following key elements: antennas, electronic
components, network architectures, protocols and signal processing.

The first important ingredient of massive MIMO networks is antenna
technology, which allows designers to assemble large antenna arrays
with various requirements in terms of spacing of elements and
geometries, reducing the number of required radio frequency (RF)
chains at the transmit and the receive ends and their implementation
costs \cite{waldschmidt,grau,chiu}. In certain scenarios and
deployments, the use of compact antennas with closely-spaced
elements will be of great importance to equip devices with a large
number of antennas but this will require techniques to mitigate the
coupling effects especially at the user terminals \cite{wallace}.
The second key area for innovation is that of electronic components
and RF chains, where the use of low-cost amplifiers with output
power in the mWatt range will play an important role. Architectures
such as the direct-conversion radio (DCR) \cite{schenk} are very
attractive due to their flexibility and ability to operate with
several different air interfaces, frequency bands and waveforms.
Existing peripherals such as large coaxial cables and power-hungry
circuits will have to be replaced with low-energy solutions.

Another key element of massive MIMO networks is the network
architecture, which will evolve from homogeneous cellular layouts to
heterogeneous architectures that include small cells and the use of
coordination between cells \cite{combes}. Since massive MIMO
technology is likely to be incorporated into cellular and local area
networks in the future, the network architecture will necessitate
special attention on how to manage the interference created
\cite{aggarwal} and measurements campaigns will be of fundamental
importance \cite{shepard}-\cite{gao}. The coordination of adjacent
cells will be necessary due to the current trend towards aggressive
reuse factors for capacity reasons, which inevitably leads to
increased levels of inter-cell interference and signalling. The need
to accommodate multiple users while keeping the interference at an
acceptable level will require significant work in scheduling and
medium-access protocols.

The last ingredient of massive MIMO networks and the main focus of
this article is signal processing. In particular, MIMO signal
processing will play a crucial role in dealing with the impairments
of the physical medium and in providing cost-effective tools for
processing information. Current state-of-the-art in MIMO signal
processing requires a computational cost for transmit and receive
processing that grows as a cubic or super-cubic function of the
number of antennas, which is clearly not scalable with a large
number of antenna elements. We advocate the need for simpler
solutions for both transmit and receive processing tasks, which will
require significant research effort in the next years. Novel signal
processing strategies will have to be developed to deal with the
problems associated with massive MIMO networks like computational
complexity and its scalability, pilot contamination effects, RF
impairments, coupling effects, delay and calibration issues. Another
key point for future massive MIMO technology is the application
scenarios, which will become the main object of investigation in the
coming years. Amongst the most important scenarios are multi-beam
satellite networks, cellular systems beyond LTE-A \cite{lte-a} and
local area networks.

%
%
%
%

This article is structured as follows. Section II reviews the system
model including both uplink and downlink and discusses the
application scenarios. Section III is dedicated to transmit
processing techniques, whereas Section IV concentrated on receive
processing. Section V discusses the results of some simulations and
Section VI presents some open problems and suggestions for further
work. The conclusions of this article are given in Section VII.

\section{Application Scenarios and Signal Models}

In this section, we discuss several application scenarios for
multiuser massive MIMO systems which include multibeam satellite
systems, cellular and local area networks. Signal models based on
elementary linear algebra are then presented to describe the
information processing in both uplink and downlink transmissions.
These models are based on the assumption of a narrowband signal
transmission over flat fading channels which can be easily
generalized to broadband signal transmission with the use of
multi-carrier systems.

\subsection{Application Scenarios}

\begin{figure}[!htb]
\begin{center}
\def\epsfsize#1#2{1\columnwidth}
\epsfbox{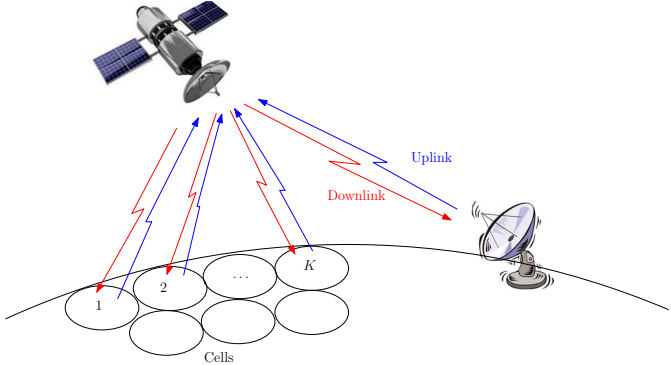} \vspace{-1em} \caption{Multi-beam satellite
network.}\label{fig1}
\end{center}
\end{figure}

Amongst the most promising application scenarios of multiuser
massive MIMO techniques are multibeam satellite \cite{arnau},
cellular and local area networks. Multibeam satellite systems are
perhaps the most natural scenario for massive MIMO because the
number of antenna elements is above one hundred. The major benefit
of satellite communications is that all users can be served within
the coverage region at the same cost. In this context, the next
generation of broadband satellite networks will employ multibeam
techniques in which the coverage region is served by multiple spot
beams intended for the users that are shaped by the antenna feeds
forming part of the payload \cite{arnau}, as depicted in Fig 1. A
fundamental problem with the multibeam approach is the interference
caused by multiple adjacent spot beams that share the same frequency
band. This interference between spot beams must be mitigated by
suitable signal processing algorithms. Specifically, multiuser
interference mitigation schemes such as precoding or multiuser
detection can be jointly designed with the beamforming process at
the gateway station. The interference mitigation must be applied to
all the radiating signals instead of the user beams directly. In the
downlink (also known as the forward link in the satellite
communications literature), the interference mitigation problem
corresponds to designing transmit processing or precoding strategies
that require the channel state information (CSI). For the uplink
(also known as the reverse link), the interference mitigation
problem can be addressed by the design of multiuser detectors.

\begin{figure}[!htb]
\begin{center}
\def\epsfsize#1#2{1\columnwidth}
\epsfbox{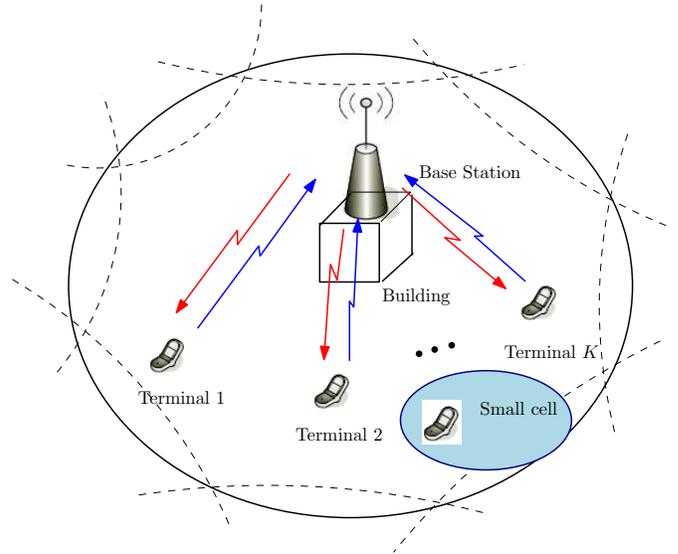} \vspace{-1em} \caption{Mobile cellular
network.}\label{fig2}
\end{center}
\end{figure}

The second highly-relevant scenario is that of mobile cellular
networks beyond LTE-A \cite{lte-a}, which is illustrated in Fig. 2.
In such networks, massive MIMO would play a key role with the
deployment of hundreds of antenna elements at the base station,
coordination between cells and a more modest number of antenna
elements at the user terminals. At the base station, very large
antenna arrays could be deployed on the roof or on the façade of
buildings. With further development in the area of compact antennas
and techniques to mitigate mutual coupling effects, it is likely
that the number of antenna elements at the user terminals (mobile
phones, tables and other gadgets) might also be significantly
increased from $1-4$ elements in current terminals to $10-20$ in
future devices. In these networks, it is preferable to employ
time-division-duplexing (TDD) mode to perform uplink channel
estimation and obtain downlink CSI by reciprocity for signal
processing at the transmit side. This operation mode will require
cost-effective calibration algorithms. Another critical requirement
is the uplink channel estimation, which employs non-orthogonal
pilots and due to the existence of adjacent cells and the coherence
time of the channel needs to reuse the pilots \cite{jose}. Pilot
contamination occurs when CSI at the base station in one cell is
affected by users from other cells. In particular, the uplink (or
multiple-access channel) will need CSI obtained by uplink channel
estimation, efficient multiuser detection and decoding algorithms.
The downlink (also known as the broadcast channel) will require CSI
obtained by reciprocity for transmit processing and the development
of cost-effective scheduling and precoding algorithms.

\begin{figure}[!htb]
\begin{center}
\def\epsfsize#1#2{0.8\columnwidth}
\epsfbox{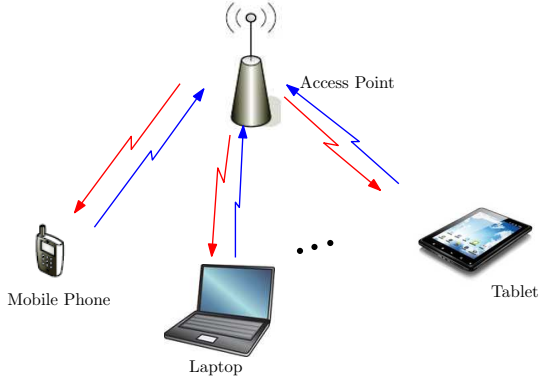} \vspace{-1em} \caption{Wireless local area
network.}\label{fig3}
\end{center}
\end{figure}

The third and last highly-relevant scenario is represented by
wireless local area networks (WLANs) \cite{wlan}, which are shown in
Fig. 3. The deployment of WLANs has increased tremendously in the
last few years with the proliferation of hot spots and home users.
These systems have adopted orthogonal frequency-division
multiplexing (OFDM) for their air interface and are equipped with a
number of antennas of up to $8$ at the access point and up to $4$
antennas at the user terminals \cite{wlan}. Massive MIMO could play
an important role in the incorporation of a substantial number of
antenna elements at the access point using compact antennas and
planar array geometries to keep the size of the access point at
reasonable physical dimensions. The user terminals (laptops, tablets
and smart phones) could also rely on compact antennas to accommodate
a substantial number of radiating elements. In the future, it is
possible that the number of antenna elements at the user terminals
will be significantly increased from $8$ to over $100$ elements at
the access points terminals and from $4$ to over $40$ in future
devices.

A key challenge in all the three scenarios is how to deal with a
very large number of antenna elements and develop cost-effective
algorithms, resulting in excellent performance in terms of the
metrics of interest, namely, bit error rate (BER), sum-rate and
throughput. In what follows, signal models that can describe the
processing and transmission will be detailed.

\subsection{Donwlink Model}

In our description, we consider a multiuser massive MIMO system with
a number of antenna elements equal to $N_A$ at the transmitter,
which could be a satellite gateway, a base station of a cellular
network or an access point of a WLAN. The transmitter communicates
with $K$ users in the system, where each user is equipped with $N_U$
antenna elements and $N_A > K N_U$. It should be noted that in
massive MIMO systems, it is desirable to have an excess of degrees
of freedom \cite{marzetta_first}, which means $N_A$ should exceed
$KN_U$ by a significant margin in order to leverage the array gain.
At each time instant $[i]$, the transmitter applies a precoder to
the $K N_U$ data vector ${\boldsymbol s}[i]$ intended for the $K$
users. The $K N_U$ data vector ${\boldsymbol s}[i]$ consists of the
stacking of the $N_U \times 1$ vectors ${\boldsymbol s}_k [i] =
\big[ s_{k,1}[i], ~s_{k,2}[i], ~ \ldots,~ s_{k,N_U}[i] \big]^T$ of
the $K$ users, where each entry is a data symbol taken from a
modulation constellation $A = \{ a_1,~a_2,~\ldots,~a_N \}$ with zero
mean and variance $\sigma_s^2$, where $(\cdot)^T$ denotes transpose.
The $N_A \times 1$ precoded data vector for user $k$ is given by
${\boldsymbol x}_k[i] = {\mathcal P}({\boldsymbol s}_k[i])$, where
${\mathcal P}(\cdot)$ is the mathematical mapping applied by the
precoder, and is then transmitted over flat fading channels.

The received signal at each user after demodulation, matched
filtering and sampling is collected in an $N_U \times 1$ vector
${\boldsymbol r}_k[i] = \big[ r_{k,1}[i], ~r_{k,2}[i], ~ \ldots,~
r_{k,N_U}[i] \big]^T$ with sufficient statistics for processing and
given by
\begin{equation}
{\boldsymbol r}_k[i] = \sum_{k=1}^{K}{\boldsymbol H}_k {\boldsymbol
x}_k[i] + {\boldsymbol n}_k[i],
\end{equation}
where the $N_U \times 1$ vector ${\boldsymbol n}_k[i]$ is a zero
mean complex circular symmetric Gaussian noise with covariance
matrix $E\big[ {\boldsymbol n}_k[i] {\boldsymbol n}^H_k[i] \big] =
\sigma_n^2 {\boldsymbol I}$, where $E[ \cdot]$ stands for expected
value, $(\cdot)^H$ denotes the Hermitian operator, $\sigma_n^2$ is
the noise variance and ${\boldsymbol I}$ is the identity matrix. The
$N_A \times 1$ precoded data vectors ${\boldsymbol x}_k[i]$ have
covariance matrices $E\big[ {\boldsymbol x}_k[i] {\boldsymbol
x}_k^H[i] \big] = \sigma_{x_{k}}^2 {\boldsymbol I}$, where
$\sigma_{x_{k}}^2$ is the signal power. The elements $h_{n_U,n_A}$
of the $N_U \times N_A$ channel matrices ${\boldsymbol H}_k$ are the
complex channel gains from the $n_A$th transmit antenna to the
$n_U$th receive antenna.

\subsection{Uplink Model}

Let us now consider the uplink of a multiuser massive MIMO system
with $K$ users that are equipped with $N_U$ antenna elements and
communicate with a receiver with $N_A$ antenna elements, where $N_A
> K N_U$. At each time instant, the $K$ users transmit
$N_U$ symbols which are organized into a $N_U \times 1$ vector
${\boldsymbol s}_k [i] = \big[ s_{k,1}[i], ~s_{k,2}[i], ~ \ldots,~
s_{k,N_U}[i] \big]^T$ taken from a modulation constellation $A = \{
a_1,~a_2,~\ldots,~a_N \}$. The data vectors ${\boldsymbol s}_k[i]$
are then transmitted over flat fading channels. The received signal
after demodulation, matched filtering and sampling is collected in
an $N_A \times 1$ vector ${\boldsymbol r}[i] = \big[ r_1[i],
~r_2[i], ~ \ldots,~ r_{N_R}[i] \big]^T$ with sufficient statistics
for processing as described by
\begin{equation}
{\boldsymbol r}[i] = \sum_{k=1}^{K}{\boldsymbol H}_k {\boldsymbol
s}_k[i] + {\boldsymbol n}[i],
\end{equation}
where the $N_A \times 1$ vector ${\boldsymbol n}[i]$ is a zero mean
complex circular symmetric Gaussian noise with covariance matrix
$E\big[ {\boldsymbol n}[i] {\boldsymbol n}^H[i] \big] = \sigma_n^2
{\boldsymbol I}$. The data vectors ${\boldsymbol s}_k[i]$ have zero
mean and covariance matrices $E\big[ {\boldsymbol s}_k[i]
{\boldsymbol s}_k^H[i] \big] = \sigma_{s_{k}}^2 {\boldsymbol I}$,
where $\sigma_{s_{k}}^2$ is the signal power. The elements
$h_{n_A,n_U}$ of the $N_A \times N_U$ channel matrices ${\boldsymbol
H}_k$ are the complex channel gains from the $n_U$th transmit
antenna to the $n_A$th receive antenna.

\section{Transmit Processing}

In this section, we discuss several aspects related to transmit
processing in massive MIMO systems.   Fundamental results in
information theory have shown that the optimum transmit strategy for
the multiuser massive MIMO downlink channel involves a theoretical
dirty paper coding (DPC) technique that performs interference
cancellation combined with an implicit user scheduling and power
loading algorithm \cite{caire}. However, this optimal approach is
extremely costly and unlikely to be used in any practical
deployment. In what follows, we consider several aspects of transmit
processing in massive MIMO systems which include TDD operation,
pilot contamination, resource allocation and precoding, and related
signal processing tasks.

\subsection{TDD operation}

One of the key problems in modern wireless systems is the
acquisition of CSI in a timely way. In time-varying channels, TDD
offers the most suitable alternative to obtain CSI because the
training requirements in a TDD system is independent of the number
of antennas at the base station (or access point) \cite{jose} and
there is no need for CSI feedback. In particular, TDD systems rely
on reciprocity by which the uplink channel is used as an estimate of
the downlink channel. An issue in this operation mode is the
difference in the transfer characteristics of the amplifiers and the
filters in the two directions. This can be addressed through
measurements and appropriate calibration \cite{rusek}. In contrast,
in a frequency division duplexing (FDD) system the training
requirements is proportional to the number of antennas and CSI
feedback is essential. For this reason, massive MIMO systems will
most likely operate in TDD mode and will require further
investigation in calibration methods.

\subsection{Pilot contamination}

The adoption of TDD mode and uplink training in massive MIMO systems
with multiple cells results in a phenomenon called pilot
contamination. In multi-cell scenarios, it is difficult to employ
orthogonal pilot sequences because the duration of the pilot
sequences depends on the number of cells and this duration is
severely limited by the channel coherence time due to mobility.
Therefore, non-orthogonal pilot sequences must be employed and this
affects the CSI employed at the transmitter. Specifically, the
channel estimate is contaminated by a linear combination of channels
of other users that share the same pilot \cite{jose}. Consequently,
the precoders and resource allocation algorithms will be highly
affected by the contaminated CSI. Strategies to control or mitigate
pilot contamination and its effects are very important for massive
MIMO networks. Possible approaches include work on optimization of
waveforms, blind channel estimation techniques, implicit training
approaches and precoding and resource allocation techniques that
take into account pilot contamination to mitigate its effects.

\subsection{Resource allocation}

Prior work on multiuser MIMO \cite{tu,dimic,shen} has shown that
resource allocation techniques are fundamental to obtain further
capacity gains. In massive MIMO this will be equally important and
will have the extra benefit of more accurate CSI. From a multiuser
information theoretic perspective, the capacity region boundary is
achieved by serving all $K$ active users simultaneously. The
resources (antennas, users and power) that should be allocated to
each user depend on the instantaneous CSI which may vary amongst
users. Since the total number of users $Q$ that could be served is
often much higher than the number of transmit antennas $N_A$, the
system needs a resource allocation algorithm to select the best set
of users according to a chosen criterion such as the sum rate or a
user target rate. The resource allocation task is then to choose a
set of users and their respective powers in order to satisfy a given
performance metric. In massive MIMO systems, the spatial signatures
of the users to be scheduled might play a fundamental role thanks to
the very large number of antennas and an excess of degrees of
freedom \cite{marzetta_first,rusek}. The multiuser diversity
\cite{tu} along with high array gains might be exploited by resource
allocation algorithm along with timely CSI. In particular, the
problem of user selection, i.e., scheduling, corresponds to a
combinatorial problem equivalent to the combination of $K$ choosing
$Q$. Hence, it is clear that the exhaustive search over all possible
combinations is computationally prohibitive when the $K$ in the
system is reasonably large, and thus cost-effective user selection
algorithms will be required. Strategies based on greedy, low-cost
and discrete optimization methods \cite{dimic,shen,tds2} are very
promising for massive MIMO networks because they could reduce the
cost of resource allocation algorithms.

\subsection{Precoding and Related Techniques}

Strategies for mitigating the multiuser interference at the transmit
side include transmit beamforming \cite{rusek} and precoding based
on linear minimum mean square error (MMSE) \cite{joham} or
zero-forcing (ZF) \cite{spencer} techniques and nonlinear approaches
such as DPC, Tomlinson-Harashima precoding (THP)
\cite{windpassinger} and vector perturbation \cite{peel}. Transmit
matched filtering (TMF) is the simplest method for processing data
at the transmit side and has been recently advocated by several
works for massive MIMO systems \cite{marzetta_first,rusek}. The
basic idea is to apply the conjugate of the channel matrix to the
data symbol vector ${\boldsymbol s}[i]$ prior to transmission as
described by
\begin{equation}
{\boldsymbol x}[i] = {\boldsymbol H}^H{\boldsymbol s}[i],
\end{equation}
where the $N_A \times K N_U$ matrix ${\boldsymbol H}$ contains the
parameters of all the channels and the $N_A \times 1$ vector
${\boldsymbol x}[i]$ represents the data processed by TMF.

Linear precoding techniques such as ZF and MMSE precoding are based
on channel inversion operations and are attractive due to their
relative simplicity for MIMO systems with a small to moderate number
of antennas. However, channel inversion based precoding requires a
higher average transmit power than other precoding algorithms
especially for ill conditioned channel matrices, which could result
in poor performance. A linear precoder applies a linear
transformations to the data symbol vector ${\boldsymbol s}[i]$ prior
to transmission as described by
\begin{equation}
{\boldsymbol x}[i] = {\boldsymbol W}_k{\boldsymbol s}_k[i] +
\sum_{l=1,l\neq k}^{K} {\boldsymbol W}_l {\boldsymbol s}_l[i],
\end{equation}
where the $N_A \times N_U$ matrix ${\boldsymbol W}_l$ contains the
parameters of the channels and the $N_U \times 1$ data symbol
vectors ${\boldsymbol s}_k[i]$ represent the data processed by the
linear precoder. The linear MMSE precoder is described by
${\boldsymbol W}_{\rm MMSE} = {\boldsymbol H}^H({\boldsymbol
H}{\boldsymbol H}^H + \gamma {\boldsymbol I})^{-1}$, where $\gamma$
is a gain factor, and the linear ZF precoder is expressed by
${\boldsymbol W}_{\rm ZF} = {\boldsymbol H}^H({\boldsymbol
H}{\boldsymbol H}^H)^{-1}$.

Block diagonalization (BD) type precoding algorithms have been
proposed in \cite{spencer,stankovic,zu} for MU-MIMO systems. The
main advantage of BD type algorithms is the sum-rate performance
that is not far from that obtained by DPC techniques and the
relative simplicity for implementation in systems with a modest
number of antennas. However, existing BD solutions are unlikely to
be used in massive MIMO systems due to the cost associated with
their implementation in antenna arrays with hundreds of elements.
This suggests that there is need for cost-effective BD type
strategies for very large antenna arrays. THP \cite{windpassinger}
is a non-linear precoding technique that employs feedforward and
feedback matrices along with a modulo operation to cancel the
multiuser interference in a more effective way than a standard
linear precoder. With THP, the $N_A \times 1$ precoded data vector
is given by
\begin{equation}
{\boldsymbol x}[i] = {\boldsymbol F} \tilde{\boldsymbol x}[i],
\end{equation}
where ${\boldsymbol F}$ is the $N_A \times K N_U$ feedforward
precoding matrix which can be obtained by an LQ decomposition of the
channel matrix ${\boldsymbol H}$ and the input data
$\tilde{\boldsymbol x}[i]$ is computed element-by-element by
\begin{equation}
\tilde{x}_l[i] = {\rm mod} \big\{ s_l[i] - \sum_{q=1}^{l-1} b_{lq}
x_q[i] \big\}, ~~ l = 1, \ldots, K N_U,
\end{equation}
where $b_{lq}$ are the elements of the $KN_U \times K N_U$ lower
triangular matrix ${\boldsymbol B}$ that can also be obtained by an
LQ decomposition. Amongst the appealing features of THP are its
excellent BER and sum-rate performances which are not far from DPC
and its flexibility to incorporate channel coding. Future work on
THP for massive MIMO networks should concentrate on the reduction of
the computational cost to compute the feedforward and feedback
matrices since existing factorization algorithms would be too costly
for systems with hundreds of antenna elements.

Vector perturbation employs a modulo operation at the transmitter to
perturb the transmitted signal vector and to avoid the transmit
power enhancement incurred by ZF or MMSE methods \cite{peel}. The
task of finding the optimal perturbation involves solving a minimum
distance type problem that can be implemented using sphere encoding
or full search-based algorithms. Let ${\boldsymbol H}$ denote a $N_A
\times K N_U$ multiuser composite channel. The idea of perturbation
is to find a perturbing vector ${\boldsymbol p}$ from an extended
constellation to minimize the transmit power. The perturbation
${\boldsymbol p}$ is obtained by solving
\begin{equation}
{\boldsymbol p}[i] = \arg \min_{{\boldsymbol p}'[i] \in A CZ^K}
||{\boldsymbol W}({\boldsymbol s}[i] + {\boldsymbol p}'[i]||^2
\end{equation}
where ${\boldsymbol W}$ is some linear transformation or precoder
such that $Tr({\boldsymbol W}^H{\boldsymbol W}) \leq P$,  the scalar
$A$ is chosen depending on the constellation size (e.g., $A = 2$ for
QPSK), and $CZ^K$ is the K-dimensional complex lattice. The transmit
matched filter, linear ZF or MMSE precoders can be used for
${\boldsymbol W}$. After pre-distortion using a linear precoder, the
resulting constellation region also becomes distorted and thus a
modulo operation is employed. This problem can be regarded as
K-dimensional integer-lattice least squares problem, which can be
solved by search based algorithms \cite{peel}.

\section{Receive Processing}

In this section, we discuss receive processing in massive MIMO
systems. In particular, we examine parameter estimation and
detection algorithms, iterative detection and decoding techniques,
mitigation of RF impairments and related signal processing tasks.

\subsection{Parameter Estimation and Detection Algorithms}

Amongst the key problems in the uplink of multiuser massive MIMO
systems are the estimation of parameters such as channels gains and
receive filter coefficients, and the detection of the transmitted
symbols ${\boldsymbol s}_k$  of each user as described by the signal
model in (2). The parameter estimation task usually relies on pilot
(or training) sequences and signal processing algorithms. In
multiuser massive MIMO networks, non-orthogonal training sequences
are likely to be used in most application scenarios and the
estimation algorithms must be able to provide the most accurate
estimates and to track the variations due to mobility. Standard MIMO
linear MMSE and least-squares (LS) channel estimation algorithms
\cite{biguesh} can be used for obtaining CSI. However, the cost
associated with these algorithms is often cubic in the number of
antenna elements at the receiver, i.e., $N_A$ in the uplink.
Moreover, in scenarios with mobility the receiver will need to
employ adaptive algorithms \cite{haykin} which can track the channel
variations. Interestingly, massive MIMO systems have an excess of
degrees of freedom that translates into a reduced-rank structure to
perform parameter estimation. This is an excellent opportunity that
massive MIMO offers to apply reduced-rank algorithms
\cite{qian}-\cite{jio_mimo} and further develop these techniques.

In order to separate the data streams transmitted by the different
users in a multiuser massive MIMO network, a designer must resort to
detection techniques, which are similar to multiuser detection
methods \cite{verdu}. The optimal maximum likelihood (ML) detector
is described by
\begin{equation}
\hat{\boldsymbol s}_{\rm ML}[i] = \arg \min_{{\boldsymbol s}[i]}
||{\boldsymbol r}[i] - {\boldsymbol H}{\boldsymbol s}[i]||^2
\end{equation}
where the $KN_U \times 1$ data vector ${\boldsymbol s}[i]$ contains
the symbols of all users. The ML detector has a cost that is
exponential in the number of data streams and the modulation order
that is too complex to be implemented in systems with a large number
of antennas. Even though the ML solution can be alternatively
computed using sphere decoder (SD) algorithms
\cite{viterbo}-\cite{shim} that are very efficient for MIMO systems
with a small number of antennas, the cost of SD algorithms depends
on the noise variance, the number of data streams to be detected and
the signal constellation, resulting in high computational costs for
low signal-to-noise ratios (SNR), high-order constellations and a
large number of data streams.

The high computational complexity of the ML detector and the SD
algorithms in the scenarios described above have motivated the
development of numerous alternative strategies for MIMO detection,
which often rely on signal processing with receive filters. The key
advantage of these approaches with receive filters is that the cost
is typically not dependent on the modulation and the receiver can
compute the receive filter only once per data packet and perform
detection. Algorithms that can compute the parameters of receive
filters with low cost are of central importance to massive MIMO
systems. In what follows, we will briefly review some relevant
suboptimal detectors, which include linear and decision-driven
strategies.

Linear detectors \cite{duel_mimo} include approaches based on the
receive matched filter (RMF), ZF and MMSE designs and are described
by
\begin{equation}
\hat{\boldsymbol s}[i] = Q\big( {\boldsymbol W}^H{\boldsymbol r}[i]
\big),
\end{equation}
where the receive filters are ${\boldsymbol W}_{\rm RMF} =
{\boldsymbol H}$ for the RMF, ${\boldsymbol W}_{\rm MMSE} =
({\boldsymbol H}{\boldsymbol H}^H + \sigma_s^2/\sigma_n^2
{\boldsymbol I})^{-1}{\boldsymbol H}$ for the MMSE and ${\boldsymbol
W}_{\rm ZF} = ({\boldsymbol H}{\boldsymbol H}^H)^{-1}{\boldsymbol
H}$ for the ZF design, and $Q(\cdot)$ represents the slicer used for
detection.

Decision-driven detection algorithms such as successive interference
cancellation (SIC) approaches used in the {Vertical-Bell
Laboratories Layered Space-Time (VBLAST)} systems
\cite{vblast}-\cite{peng_twc} and decision feedback (DF) \cite{choi}
detectors are techniques that can offer attractive trade-offs
between performance and complexity. Prior work on SIC and DF schemes
has been reported with DF detectors with SIC (S-DF)
\cite{choi,varanasi3} and DF receivers with parallel interference
cancellation (PIC) (P-DF) \cite{woodward2,delamare_mber},
combinations of these schemes \cite{woodward2,spa,mdfpic} and
mechanisms to mitigate error propagation
\cite{reuter,delamare_itic}. DF detectors
\cite{choi,varanasi3,woodward2} employ feedforward and feedback
matrices that can be based on the receive matched filter (RMF), ZF
and MMSE designs as described by
\begin{equation}
\hat{\boldsymbol s} = Q\big( {\boldsymbol W}^H{\boldsymbol r}[i] -
{\boldsymbol F}^H\hat{\boldsymbol s}_o[i] \big),
\end{equation}
where $\hat{\boldsymbol s}_o$ corresponds to the initial decision
vector that is usually performed by the linear section of the DF
receiver (e.g., $\hat{\boldsymbol s}_o = Q( {\boldsymbol
W}^H{\boldsymbol r})$) prior to the application of the feedback
section. The receive filters ${\boldsymbol W}$ and ${\boldsymbol F}$
can be computed using design criteria and optimization algorithms.

An often criticized aspect of these sub-optimal schemes is that they
typically do not achieve the full receive-diversity order of the ML
algorithm. This led to the investigation of detection strategies
such as lattice-reduction (LR) schemes
\cite{windpassinger}-\cite{gan}, QR decomposition, M-algorithm
(QRD-M) detectors \cite{kim_qrdm}, probabilistic data association
(PDA) \cite{jia,syang} and multi-branch \cite{spa,mbdf} detectors,
which can approach the ML performance at an acceptable cost for
small to moderate systems. The development of cost-effective
detection algorithms for massive MIMO systems is a formidable task
that calls for new approaches and ideas in this exciting area.

\subsection{Iterative Detection and Decoding Techniques}

Iterative detection and decoding (IDD) schemes have received
considerable attention in the last years following the discovery of
Turbo codes \cite{berrou} and the use of the Turbo principle for
mitigation of several sources of interference
\cite{douillard}-\cite{choi2}. More recently, work on IDD schemes
has been extended to low-density parity-check codes (LDPC)
\cite{hou,wu} and their variants which rival Turbo codes in terms of
performance. The basic idea of an IDD system is to combine an
efficient soft-input soft-output (SISO) detection algorithm and a
SISO decoding technique. In particular, the detector produces
log-likelihood ratios (LLRs) associated with the encoded bits and
these LLRs serve as input to the decoder. Then, in the second phase
of the detection/decoding iteration, the decoder generates a
posteriori probabilities (APPs) after a number of (inner) decoding
iterations for encoded bits of each data stream. These APPs are fed
to the detector to help in the next iterations between the detector
and the decoder, which are called outer iterations. The joint
process of detection/decoding is then repeated in an iterative
manner until the maximum number of (inner and outer) iterations is
reached. In massive MIMO systems, it is likely that either Turbo or
LDPC codes will be adopted in IDD schemes for mitigation of
multiuser, multipath, intercell and other sources of interference.
LDPC codes exhibit some advantages over Turbo codes that include
simpler decoding and implementation issues. However, LDPC codes
often require a higher number of decoding iterations which translate
into delays or increased complexity. The development of IDD schemes
and decoding algorithms that perform message passing with reduced
delays \cite{wainwright}-\cite{jingjing} are of paramount importance
in future wireless systems.

\subsection{Mitigation of RF Impairments}

The large antenna arrays used in massive MIMO systems will pose
several issues to system designers such as coupling effects,
in-phase/quadrature (I/Q) imbalances \cite{hakkarainen}, and
failures of antenna elements, which will need to be addressed. The
first potential major impairment in massive MIMO systems is due to
reduced spacing between antenna elements which result in coupling
effects. In fact, for compact antenna arrays a reduction of the
physical size of the array inevitably leads to reduced spacing
between antenna elements, which can severely reduce the multiplexing
gain. In order to address these coupling effects, receive processing
approaches will have to work with transmit processing techniques to
undo the coupling induced by the relatively close spacing of
radiating elements in the array. Another major impairment in massive
MIMO systems is I/Q imbalances in the RF chains of the large arrays.
This problem can be addressed by receive or transmit processing
techniques and require modelling of the impairments for subsequent
mitigation. When working with large antenna arrays, a problem that
might also occur is the failure of some antenna elements. Such
sensor failures are responsible for a reduction in the degrees of
freedom of the array and must be dealt by signal processing
algorithms.

\section{Simulation Results}

In this section, we illustrate some of the techniques outlined in
this article using massive MIMO configurations, namely, a very large
antenna array, an excess of degrees of freedom provided by the array
and a large number of users with multiple antennas. We consider QPSK
modulation and channels that are fixed during a data packet and that
are modeled by complex Gaussian random variables with zero mean and
variance equal to unity. The signal-to-noise ratio (SNR) in dB is
defined as $\textrm{SNR} = 10 \log_{10} \frac{N_T
\sigma_s^2}{\sigma^2}$, where $\sigma_s^2$ is the variance of the
symbols, $\sigma_n^2$ is the noise variance, and we consider data
packets of $1000$ QPSK symbols.

\begin{figure}[!htb]
\begin{center}
\def\epsfsize#1#2{1\columnwidth}
\epsfbox{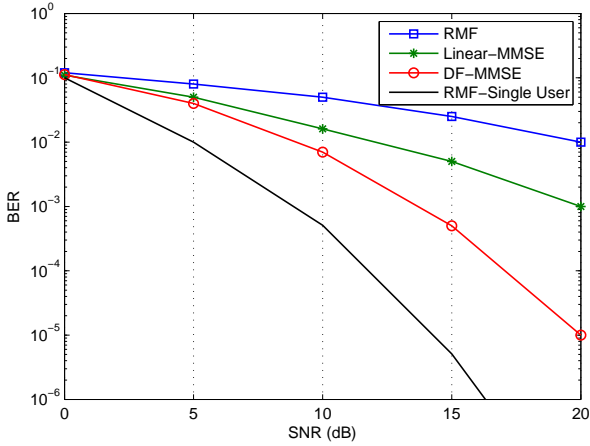} \vspace*{-0.5em} \caption{BER performance against
SNR of detection algorithms in a scenario with $N_A=128$, $K=8$
users and $N_U=8$ antenna elements.} \label{ber}
\end{center}
\end{figure}

In the first example, we compare the BER performance against the SNR
of several detection algorithms, namely, the RMF with $K=8$ users
and with a single user, the linear MMSE detector \cite{duel_mimo}
and the DF MMSE detector using a successive interference
cancellation \cite{choi,woodward2,spa}. In particular, a scenario
with $N_A=128$ antenna elements at the receiver, $K=8$ users and
$N_U=8$ antenna elements at the user devices is considered, which
corresponds to an excess of degrees of freedom equal to $N_A-K
N_U=64$. The results shown in Fig. \ref{ber} indicate that the RMF
with a single user has the best performance, followed by the DF
MMSE, the linear MMSE and the RMF detectors. Unlike previous works
\cite{rusek} that advocate the use of the RMF, it is clear that the
BER performance loss experienced by the RMF should be avoided and
more advanced receivers should be considered. However, the cost of
linear and DF receivers is dictated by the matrix inversion of $N_A
\times N_A$ matrices which must be reduced for large systems.

\begin{figure}[!htb]
\begin{center}
\def\epsfsize#1#2{1\columnwidth}
\epsfbox{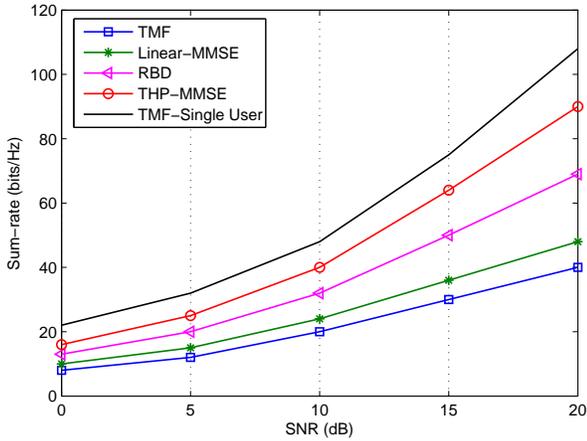} \vspace*{-0.5em} \caption{Sum-rate performance
against SNR of precoding algorithms in a scenario with $N_A=128$,
$K=8$ users and $N_U=8$ antenna elements.} \label{sr}
\end{center}
\end{figure}

In the second example, we compare the sum-rate performance against
the SNR of several precoding algorithms, namely, the TMF with a
varying number of users and with a single user, the linear MMSE
precoder and the THP MMSE precoder. The sum-rate is calculated using
\cite{vishwanath}:
\begin{equation}
C = \log(\det({\boldsymbol I} + \sigma^{-2}_n {\boldsymbol
H}{\boldsymbol P}{\boldsymbol P}^H{\boldsymbol H}^H)) {\rm
(bits/Hz)}.
\end{equation}
We consider a similar scenario to the previous one in which the
transmitter is equipped with $N_A=128$ antenna elements, and there
are $K=8$ users with $N_U=8$ antenna elements. The results in Fig.
\ref{sr} show that the TMF with a single user has the best sum-rate
performance, followed by the THP MMSE, the regularized BD (RBD), the
linear MMSE and the TMF precoding algorithms. From the curves in
Fig. \ref{sr}, we can notice that the performance of TMF is much
worse than that of THP and of RBD . This suggests that more
sophisticated precoding techniques with lower complexity should be
developed to maximize the capacity of massive MIMO systems.

\section{Future Trends and Emerging Topics}

In this section, we discuss some future signal processing trends in
the area of massive MIMO systems and point out some emerging topics
that might attract the interest of researchers. The topics are
structured as:

\begin{itemize}
\item{Transmit processing:\\

$\rightarrow$ Cost-effective scheduling algorithms: The development
of methods that have low cost and are scalable such as greedy
algorithms \cite{dimic} and discrete optimization techniques
\cite{tds2} will play a crucial role in massive MIMO networks. \\

$\rightarrow$ Calibration procedures: The transfer characteristics
of the filters and amplifiers used for TDD operation will require
designers to devise algorithms that can efficiently calibrate the links. \\

$\rightarrow$ Precoders with scability in terms of complexity: The
use of divide-and-conquer approaches, methods based on sensor array
signal processing and sectorization will play an important role to
reduce the dimensionality of the transmit processing problem.
Moreover, the investigation and development of TMF strategies with
non-linear cancellation strategies and low-cost decompositions for
linear and non-linear precoders will be important
to obtain efficient transmit methods.}\\

\item{Receive processing:\\

$\rightarrow$ Cost-effective detection algorithms: Techiques to
perform dimensionality reduction \cite{qian}-\cite{jio_mimo} for
detection problems will play an important role in massive MIMO
devices. By reducing the number of effective processing elements,
detection algorithms could be applied. In addition, the development
of schemes based on RMF with non-linear interference cancellation
capabilities might be a promising option that can close the gap
between RMF and more costly detectors.\\

$\rightarrow$ Decoding strategies with low delay: The development of
decoding strategies with reduced delay will play a key role in
applications such as audio and video streaming because of their
delay sensitivity. Therefore, we argue that novel message passing
algorithms with smarter strategies to exchange information should be
investigated along with their application to IDD schemes. \\

$\rightarrow$ Mitigation of impairments: The identification of
impairments originated in the RF chains of massive MIMO systems will
need mitigation by smart signal processing algorithms. For example,
I/Q imbalance might be dealt with using widely-linear signal
processing algorithms \cite{adali} and \cite{song}. }

\end{itemize}

\section{Concluding Remarks}

This article has presented a tutorial on massive MIMO systems and
discussed signal processing challenges and future trends in this
exciting reseach topic. Key application scenarios which include
multibeam satellite, cellular and local area networks have been
examined along with several operational requirements of massive MIMO
networks. Transmit and receive processing tasks have been discussed
and fundamental signal processing needs for future massive MIMO
networks have been identified. Numerical results have illustrated
some of the discussions on transmit and receive processing functions
and future trends have been highlighted. Massive MIMO technology is
likely to be incorporated into the applications detailed in this
article on a gradual basis by the increase in the number of antenna
elements and by the need for more sophistical signal processing
tools to transmit and process a large amount of information.

\end{document}